\documentstyle[epsfig,12pt,preprint,tighten,aps]{revtex}
\begin{document}
\begin{titlepage}

\rightline{{\large \tt Feb 2002}}

\vskip 1.4 cm

\centerline{\Large \bf
Exotic meteoritic phenomena: 
}
\vskip 0.12cm
\centerline{\Large \bf
The Tunguska event 
and anomalous low
}
\vskip 0.12cm
\centerline{\Large \bf
altitude fireballs --
manifestations of the mirror world?
}

\vskip 1.2 cm

\centerline{\large R. Foot and T. L. Yoon
}
\vskip 0.7 cm\noindent
\centerline{{\large \it foot, tlyoon@physics.unimelb.edu.au}}

\centerline{{\large \it School of Physics}}

\centerline{{\large \it Research Centre for High Energy Physics}}

\centerline{{\large \it The University of Melbourne}}

\centerline{{\large \it Victoria
 3010 Australia }}

\vskip 1.4cm

\centerline{\large Abstract} \vskip 0.5cm \noindent

There are a number of very puzzling meteoritic events including
(a) The Tunguska event. It is the only known example of a
low altitude atmospheric explosion. It is also the
largest recorded event. Remarkably no fragments or significant chemical
traces have ever been recovered. (b) Anomalous low altitude fireballs which
(in some cases) have been observed to hit the ground.
The absence of fragments is particularly striking
in these cases, but this is not the only reason they
are anomalous. The other main puzzling feature is the lack
of a consistent trajectory: low altitude fireballs, if
caused by an ordinary cosmic body penetrating the Earth's atmosphere,
should have been extremely luminous at high altitudes. But
in these anomalous
cases this is (remarkably) not observed to occur!
%Finally, we have (c) light from speedy meteors beginning at
%anomalously high altitudes -- too high to be easily
%explained by conventional mechanisms.
On the other hand, there is strong evidence that most
of our galaxy is made from exotic dark material --
`dark matter'. Mirror matter is one well motivated dark
matter candidate, since it is dark and stable and it is required to exist
if particle interactions are mirror symmetric.
If mirror matter is the dark matter, then some amount must exist
in our solar system. Although there is not much room for
a large amount of mirror matter in the inner solar system,
numerous small asteroid sized mirror matter objects
are a fascinating possibility because they can potentially
collide with the Earth.
We demonstrate that the mirror matter theory allows for a
simple explanation for the puzzling meteoritic events [both (a) and (b)]
if they are due to mirror matter space-bodies. A direct
consequence of this explanation is that mirror matter
fragments should exist in (or on) the ground 
at various impact sites.
The properties of this potentially recoverable material
depend importantly on the sign of the photon-mirror
photon kinetic mixing parameter, $\epsilon$.
We argue that the broad characteristics of the anomalous
events suggests that $\epsilon$ is probably negative.
Strategies for detecting mirror matter in the ground are discussed.

\end{titlepage}
\section{Introduction}
\vskip 1.0cm

One of the most natural candidates for a symmetry of nature is
parity symmetry (also called left-right or mirror symmetry). While
it is an established experimental fact that parity symmetry
appears broken by the interactions of the known elementary
particles, this however does not exclude the possible existence of
exact unbroken parity symmetry in nature. This is because parity
(and also time reversal) can be exactly conserved if a set of
mirror particles exist\cite{ly,flv}.  The idea is that for each
ordinary particle, such as the photon, electron, proton and
neutron, there is a corresponding mirror particle, of exactly the
same mass as the ordinary particle\footnote{
The mirror particles only have the same mass as 
their ordinary counterparts
provided that the mirror symmetry is unbroken.
It is possible to write down gauge models where
the mirror symmetry is broken\cite{broken,ug}, in some cases
allowing the mirror particles to have completely
arbitrary masses\cite{ug}, however these scenarios
tend to be more complicated and much less well motivated 
in our view.}.
Furthermore, the mirror particles interact with each other in
exactly the same way that the ordinary particles do. It follows that
the mirror proton
is stable for the same reason that the ordinary proton is stable,
and that is, the interactions of the mirror particles conserve a
mirror baryon number. The mirror particles are not produced
(significantly) in laboratory experiments just because they couple
very weakly to the ordinary particles. In the modern language of
gauge theories, the mirror particles are all singlets under the
standard $G \equiv SU(3)\otimes SU(2)_L \otimes U(1)_Y$ gauge
interactions. Instead the mirror fermions interact with a set of
mirror gauge particles, so that the gauge symmetry of the theory
is doubled, i.e. $G \otimes G$ (the ordinary particles are, of
course, singlets under the mirror gauge symmetry)\cite{flv}.
Parity is conserved because the mirror fermions experience $V+A$
(right-handed) mirror weak interactions and the ordinary fermions
experience the usual $V-A$ (left-handed) weak interactions.
Ordinary and mirror particles interact with each other
predominately by gravity only.

At the present time there is a large range of experimental
observations supporting the existence of mirror matter, for a
review see Ref.\cite{puz} (for a more detailed discussion
of the case for mirror matter, accessible to
the non-specialist, see the recent book\cite{bk}).  
The evidence includes numerous observations
suggesting the existence of invisible `dark matter' in galaxies.
Mirror matter is stable and dark and provides a natural candidate
for this inferred dark matter\cite{blin}. 
The MACHO observations\cite{mo}, close-in extrasolar
planets\cite{ce}, isolated planets\cite{is} and
even gamma ray bursts\cite{grb} may
all be mirror world manifestations.
On the quantum level, small fundamental
interactions connecting ordinary and mirror matter are possible.
Theoretical constraints from gauge invariance, renormalizability
and mirror symmetry suggest only three possible types of
interactions\cite{flv,flv2}: photon-mirror photon kinetic mixing,
neutrino-mirror neutrino mass mixing and Higgs-mirror Higgs
interactions. The main experimental implication of photon-mirror
photon kinetic mixing is that it modifies the properties of
orthopositronium, leading to a shorter effective lifetime in
`vacuum' experiments\cite{gl,gn,fg}. A shorter lifetime is in fact
seen at the 5 sigma level!\cite{vac,fg}. Neutrino-mirror neutrino
mass mixing implies maximal oscillations for each ordinary
neutrino with its mirror partner\cite{flv2,mm} -- a result
which may be connected with the neutrino physics anomalies\cite{zz5}. 

The purpose of the present paper is to make a detailed study of
one very explosive implication of the mirror matter theory, and
that is, that our solar system contains small asteroid sized
mirror matter space bodies which occasionally collide with our planet. In
Ref.\cite{tunguska,puz} it was proposed that such mirror matter space
bodies may have caused the famous 1908 Siberian explosion -- the
Tunguska event -- as well as other smaller, but more frequent
events. In the present paper we will examine this idea in more
detail. We will show that the mirror matter space-body (SB)
hypothesis provides a natural framework for a unified explanation
for a number of puzzling meteoritic events which do not seem to be
naturally associated with an ordinary matter SB, including the
1908 Tunguska event and the anomalous low altitude fireball events. 

\section{Some puzzling observations}

Our solar system contains a large variety of small
space bodies (SB) -- asteroids and comets -- as well as the
9 known planets and the various moons.
Although tiny, small SB may be very numerous
and may have big implications for life on our planet.
The reason is that sometimes they might collide with our
planet releasing large amounts of energy in the process.
For example, there is interesting evidence
that the mass extinction which wiped out
the dinosaurs 65 million years ago was caused by
the collision of a large asteroid or comet with the Earth.
The evidence is in the form of an excess of the
rare element iridium in clay samples dating
from that time period\cite{alvarez}. Iridium is very rare in
the Earth's crust and mantle but much more
common in asteroids and comets. There is also evidence
for a large meteorite crater also dating from
the same time period. It is located in the
Yucatan peninsula of Mexico. The estimated
size of this asteroid is of order 10 kilometers
in diameter with a mass of about 500 billion tons.

More recently, there is evidence that
an object of order 50 metres in size
collided with the Earth in 1908 causing a very large
explosion in the Tunguska river region of Siberia.
However, while the impact 65 million years ago left
chemical traces (the excess of iridium) as well as a
crater, the more recent Tunguska
object is somewhat more inconspicuous - and much more
puzzling.

\subsection{The Tunguska event}

In the early morning of June 30th 1908 a powerful explosion
occurred in the Tunguska river region of Siberia.
The explosion flattened about 2,100 square kilometers of forest
in a radial pattern (see Figure 1). 
The energy released in the explosion has been estimated to be
the equivalent of roughly 20 megatons of TNT or 1000 atomic bombs.
There was also evidence that the inner two hundred square kilometers
of trees was burned from above.  
%\linebreak
\newpage

\vskip 0.1cm
%\begin{figure}[t]
%\begin{center}
%\epsfig{file=fig1a.eps,height = 12.9cm, width=11cm}

.

\vskip 1.0cm

--------------------------------------------------------
-------------------------------------------------------

\vskip 1cm

\begin{center}
Figure 1a (here)

\vskip 1cm
--------------------------------------------------------
-------------------------------------------------------
\vskip 3.5cm
\epsfig{file=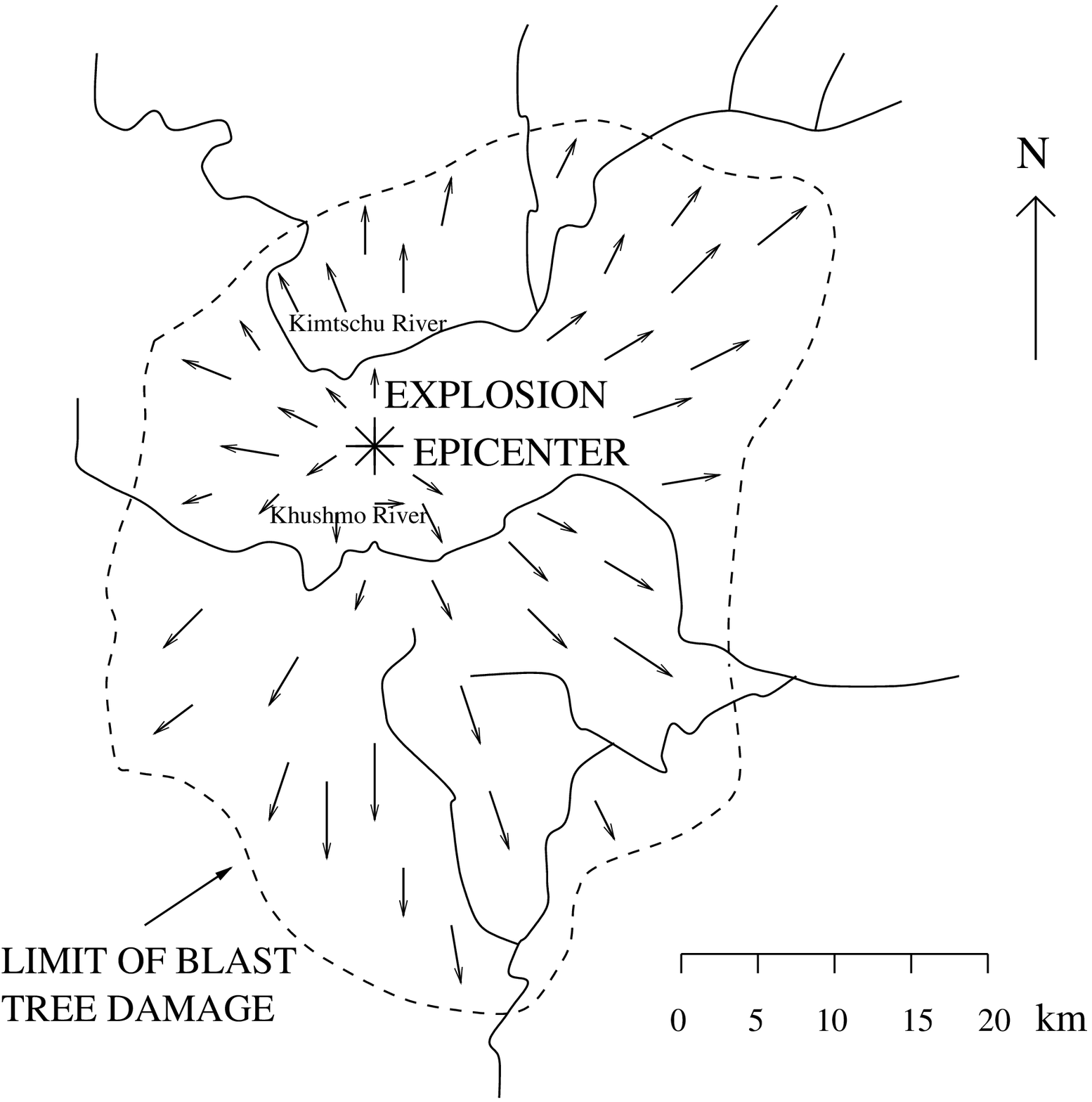,width=10.5cm}
\end{center}
\vskip 0.6cm
\noindent
Fig. 1: The forest devastation at Tunguska.
The top figure shows the fallen trees on the banks of
the Khushmo river as seen by Kulik in 1928. The bottom figure
shows the area and orientation of the fallen trees.
\newpage

\noindent
The broad features of the event
suggest a huge explosion in the atmosphere at an
altitude of between about 5-8 km which produced a downward
going spherical shockwave. The spherical 
shockwave toppled the
trees in the radial pattern and the heat from the
explosion caused the 
flash burn of the trees.
For a recent review
of what is known about the Tunguska event, see Ref.\cite{vasilyev}.

It is a remarkable fact that after considerable experimental study
with more than 40 scientific expeditions to the site, the origin
of the Tunguska explosion is still an open question.  To explain the
forest fall and other features 
requires a relatively low altitude explosion ($\sim
5-8 \ $km height), which suggests that the cosmic body was able to
withstand huge pressures without breaking up or completely
ablating. Roughly, an ordinary body should break up when the
pressure at its surface exceeds its mechanical strength.
Furthermore, a large body, like the Tunguska body, would not lose
much of its cosmic velocity during its atmospheric flight while it
remains intact. Thus, as the body moves closer to the Earth's
surface the  pressure quickly increases in proportion to the
increasing density of the Earth's atmosphere.  It has been argued
that the necessary low altitude of the explosion, indicated by the
broad features of the forest fall, suggests that the body should be
mechanically strong of asteroidal composition rather than
cometary. However, the break up of a mechanically strong body made
of non-volatile material may be expected to lead to multiple
explosions and macroscopic fragments (as well as significant
chemical traces, such as iridium excess) covering the `impact
region'. Yet, the evidence suggests a single predominant
explosion. Furthermore, while there is evidence for subsequent
explosions these were very small, and seem to be at much lower
altitude. In the words of Vasilyev\cite{vas}: \vskip 0.1cm
\noindent `We may tentatively conclude that along with a great
energy release from 5 to 8 kilometers above the Earth, there were
a number of low-altitude (maybe even right above the surface)
explosions that contributed to the total picture of destruction.

...It should be emphasised that though the patchiness of the
effects associated with the Tunguska explosion has been
noted in the literature more than once, its origin has not
been discussed. This seems to be due to serious difficulties of
its interpretation in terms of the existing Tunguska cosmic
body models.'

\vskip 0.2cm

On the other hand, the lack of remnants could point
to a body made of volatile material such as ices,
which could have completely vaporized in the atmosphere.
However, such a body should not have survived to low
altitudes before breaking up,
especially since comets should impact with relatively
high velocities ($v \stackrel{>}{\sim} 30$ km/s) because
of their elliptical orbits.
While it is believed that ices are the
main components of comets, it is also known that comets
typically contain significant amounts of non-volatile materials
as well\footnote{
The puzzling nature of the Tunguska event has also led
to suggestions that its origin 
was purely geophysical (see for example,
Ref.\cite{andrei}). Given the lack of
direct material evidence for the standard extraterrestrial
explanation (i.e. asteroid or comet), such alternative explanations
are interesting and possible.  However, there were numerous eye 
witness reports 
observing the large fireball heading towards
Tunguska. It is also true that some details of these reports
were contradictory, they nevertheless do support
an extraterrestrial explanation for the event (in our opinion).}.
Thus, a cometary origin of the Tunguska cosmic body cannot
really explain the lack of fragments and chemical traces. 
In either case, the evidence for lower altitude secondary
explosions does suggest that significant pieces of the
original body survived the main explosion -- but where
are the traces?

It is an interesting observational fact that, on smaller scales,
there do not seem to be events which exactly mimic the Tunguska
example. It is the only known case of a cosmic body exploding at
low altitudes in the atmosphere. Yet, there are very puzzling examples
of small bodies which have been apparently observed to survive to low
altitudes and strike the ground. In a sense they are
`Tunguska-like' because of their lack of fragments and chemical
traces (which is even more mysterious because the small bodies
have lost their cosmic velocity and strike the ground with
relatively low velocities of order 1 km/s). We will discuss some
examples of these rather mysterious impact events in part B below.
More generally fireballs disintegrate or explode at high altitudes
($\stackrel{>}{\sim} 30 $ km). An example of a high altitude
explosion (or `airburst') is given by the Lugo fireball\cite{luigi}.

On January 19, 1993 a bright fireball crossed the sky of northern
Italy, ending with an explosion roughly over the town of Lugo. The
energy of the explosion -- estimated to be about 14 thousand tons
of TNT or one atomic bomb -- generated shock waves which were
recorded by six local seismic stations. By means of the seismic
data, it was possible to calculate the height of the explosion,
which was estimated to be approximately 30 km. No fragments were
recovered. This event appears to be similar to the Tunguska event,
but with about 1000 times smaller in energy release and also the
explosion occurred at significantly higher altitude ($30 \ $km
rather than $\sim 5\ $km). Literally hundred's of other airburst
events have been recorded by the US department of Defense
satellite system (with energies in the range of 1 - 100 thousand
tons of TNT). Interestingly, they all appear to airburst
at high altitudes. The Tunguska explosion appears to be unique for
two reasons: It is the {\it largest} recorded atmospheric
explosion and also the {\it only} known example of a low altitude
airburst.

\subsection{Some examples of anomalous small fireballs}

There are many reported examples of atmospheric
phenomena resembling fireballs, which cannot be
due to the penetration of an ordinary meteoroid into the
atmosphere (for a review of bolides, including discussion
of these anomalous events, see Ref.\cite{bol}).
Below we discuss several examples of this strange
class of phenomena.

\vskip 0.2cm
\noindent
(i) {\it The Spanish event -- January 18, 1994.}

\vskip 0.2cm

On the early morning of 1994 January 18, a very bright luminous
object crossed the sky of Santiago de Compostela, Spain. This
event has been investigated in detail in Ref.\cite{docobo}. The
eye witnesses observed the object to be low in altitude and
velocity ($1$ to $3$ km/s). Yet, an ordinary body penetrating
deep into the atmosphere should have been quite large and luminous
when it first entered the atmosphere at high altitudes with large
cosmic velocity (between $11$ and $70$  km/s). An ordinary body
entering the Earth's atmosphere at these velocities always
undergoes significant ablation as the surface of the body melts
and vapourises, leading to a rapid diminishing of the bodies size
and also high luminosity as the ablated material is heated to high
temperature as it dumps its kinetic energy into the surrounding
atmosphere. Such a large luminous object would have an estimated
brightness which would supersede the brightness of the Sun,
observable at distances of at least $500 \ $km\cite{docobo}. Sound
phenomena consisting of sonic booms should also have
occurred\cite{docobo}. Remarkably neither of these two expected
phenomena were observed for this event. The authors of
Ref.\cite{docobo} concluded that the object could {\it not} be a
meteoric fireball.

In addition, within a kilometer of the projected end point of the
``object's'' trajectory a ``crater'' was later
discovered\cite{docobo}. The ``crater'' had dimensions 
$29$ m $\times 13$ m and $1.5$ m deep. At the crater site, full-grown
pine trees were thrown downhill over a nearby road. Unfortunately,
due to a faulty telephone line on the $17^{th}$ and $18^{th}$ of
January (the fireball was seen on the $18^{th}$) the seismic
sensor at the nearby geophysical observatory of Santiago de Compostela
was inoperative at the crucial time.  After a careful
investigation, the authors of Ref.\cite{docobo} concluded that the
crater was most likely associated with the fireball event, but
could not definitely exclude the possibility of a landslide.

No meteorite fragments or any other unusual material was
discovered at the crater site.

\vskip 0.2cm
\noindent
(ii) {\it The Jordan event -- April 18, 2001.}

\vskip 0.2cm

On Wednesday $18^{th}$ April
2001, more than 100 people attending a funeral
procession saw a low altitude and low velocity fireball.
In fact, the object was observed to break up into two
pieces and each piece was observed to hit the ground.
The two impact sites were later examined by members
of the Jordan Astronomical Society.
The impact sites showed evidence of energy release
(broken tree, half burnt tree, sheared rocks and
burnt ground) but no ordinary crater (see figure 2). 
[This may have been
due, in part, to the hardness of the ground at the impact
sites].  No meteorite fragments were recovered despite
the highly localized nature of the impact sites and low
velocity of impact. For more of the
remarkable pictures and more details, see the Jordan
Astronomical Society's report\cite{jas}.
As with the 1994 Spanish event (i),
the body was apparently not observed by anyone when it 
was at high altitudes where it should have been very bright.
Overall, this
event seems to be broadly similar to the 1994 spanish
event (i). For the same reasons discussed
in (i) (above) it could not be due to an ordinary meteoric fireball.
\vskip 0.2cm
\noindent
(iii) {\it The Poland event -- January 14, 1993.}
\vskip 0.2cm
Another anomalous event, similar to the Spanish and Jordan
cases was observed in Poland, January 14, 1993\cite{mor,bol}.
Again, a low altitude, low velocity ($v \sim 1 \ $km/s) body was observed.
In this particular case there was evidence of an
enormous electrical discharge at the `impact site', which destroyed most
of the electrical appliances in nearby houses.

\vskip 0.3cm

There are many other similar examples, some of which
have been described by Ol'khovatov in Ref.\cite{andrei}. 

\newpage

%\newpage
\vskip 0.4cm
.
\vskip 2cm
------------------------------------------------------
------------------------------------------------------
\vskip 2cm
\begin{center}
Figure 2a
\vskip 2cm
------------------------------------------------------
------------------------------------------------------
\vskip 2cm
Figure 2b
\vskip 2cm
------------------------------------------------------
------------------------------------------------------
%\vskip -17.4cm
%\epsfig{file=fig2b.eps,width=13.2cm}
\end{center}
\vskip 2.6cm
\noindent
Some pictures of the impact sites (Courtesy of the
Jordan Astronomical Society[25]).
\vskip 0.9cm
\subsection{Other anomalous events - Speedy meteors}

In standard theory, light produced by a meteoroid during its
interaction with the Earth's atmosphere is caused by the ablation
process: The surface of the meteoroid melts and vapourises due to
the extreme heating of its surface by the interactions with the
atmosphere, 
leading to emission lines as the atoms in the
surrounding vapour de-excite.  However, observations\cite{speedy}
of the Leonid meteors have shown that radiation from these
extremely speedy meteors (entering the Earth's atmosphere at about
$71$ km/s) starts at an extremely high 
altitude, up to $200$ km in height. At these high 
altitudes the atmosphere is so sparse that 
the ablation process should not be occurring at all: there is
simply not enough air molecules 
to heat and evaporate an entering
meteoroid -- yet radiation exists because it is observed in rather
great detail\cite{speedy}.

Clearly, the observations (a) and (b) [and maybe even (c)] 
indicate that there are many strange
happenings a foot. These largely unexplained phenomena
do provide motivation to examine the
fantastic possibility that they may be manifestations of the
mirror world.

\section{The interactions of a mirror matter space-body
with the atmosphere}

There is not much room for a large amount of mirror matter in our
solar system. For example, the amount of mirror matter within the
Earth has been constrained to be less than $10^{-3}
M_{Earth}$\cite{sashaV}. However, we don't know enough about the
formation of the solar system to be able to exclude the existence
of a large number of  Space Bodies (SB) made of mirror matter if
they are small like comets and asteroids. The total mass of
asteroids in the asteroid belt is estimated to be only about
0.05\% of the mass of the Earth. A similar or even greater number
of mirror bodies, perhaps orbiting in a different plane or even
spherically distributed like the Oort cloud is a fascinating and
potentially explosive possibility\footnote{ Large planetary sized
bodies are also possible if they are in distant
orbits\cite{silnem} or masquerade as ordinary planets or moons
by accreting ordinary matter onto
their surfaces\cite{bk}.} if they collide with the Earth. The possibility
that such collisions occur and may be responsible for the 1908
Siberian explosion (Tunguska event) has been speculated in
Ref.\cite{puz,tunguska}. The purpose of this paper is to study
this fascinating possibility in detail.

If such small mirror bodies exist in our
solar system and happen to
collide with the Earth, what would be the consequences?
If the only force connecting mirror matter with
ordinary matter is gravity, then the consequences
would be minimal. The mirror SB would simply
pass through the Earth and nobody would know about it
unless it was so heavy as to gravitationally affect
the motion of the Earth.
While we know that ordinary and mirror
matter do not interact with each other via any
of the {\it known} non-gravitational forces,
it is possible that new interactions exist which
couple the two sectors together.
In Ref.\cite{flv,flv2}, all such interactions consistent
with gauge invariance, mirror symmetry and renormalizability
were identified, namely,
photon-mirror photon kinetic mixing, Higgs-mirror Higgs
interactions and via ordinary neutrino-mirror neutrino
mass mixing (if neutrinos have mass).
While Higgs-mirror Higgs
interactions will be tested if or when the Higgs particle
is discovered, there is currently strong evidence
for photon-mirror photon kinetic mixing\cite{fg} and also
ordinary neutrino-mirror neutrino mass mixing\cite{flv2,mm}.
Of most importance though for this paper is
the photon-mirror photon kinetic mixing interaction.

In field theory,
photon-mirror photon kinetic mixing
is described by the interaction
\begin{equation}
{\cal L} = {\epsilon \over 2}F^{\mu \nu} F'_{\mu \nu},
\label{ek}
\end{equation}
where $F^{\mu \nu}$ ($F'_{\mu \nu}$) is the field strength tensor
for electromagnetism (mirror electromagnetism). This type of
Lagrangian term is gauge invariant and renormalizable and can
exist at tree level\cite{flv,fh} or may be induced radiatively in
models without $U(1)$ gauge symmetries (such as grand unified
theories)\cite{gl,bob,cf}. One effect of ordinary photon-mirror
photon kinetic mixing is to give the mirror charged particles a
small electric charge\cite{flv,gl,bob}. That is, they couple to
ordinary photons with electric charge $\epsilon e$.

The most important experimental implication of photon-mirror
photon kinetic mixing is that it modifies the properties
of orthopositronium\cite{gl}. This effect arises due to
radiative off-diagonal contributions to the
orthopositronium, mirror orthopositronium
mass matrix. This means that orthopositronium oscillates
into its mirror partner. Decays of mirror
orthopositronium are not detected
experimentally which effectively increases the observed
decay rate\cite{gl}. Because collisions of orthopositronium destroy
the quantum coherence, this mirror world effect is most
important for experiments which are designed such that the
collision rate of the orthopositronium is low\cite{gn}.
The only accurate experiment sensitive to the mirror
world effect is the Ann Arbour vacuum cavity experiment\cite{vac}.
This experiment obtained a decay rate of
$\Gamma_{oPs} = 7.0482 \pm 0.0016 \ \mu s^{-1}$.
Normalizing this measured value with the recent
theoretical value of $7.0399 \ \mu s^{-1}$ \cite{theory} gives
\begin{equation}
{\Gamma_{oPs}(exp) \over \Gamma_{oPs}(theory)} =
1.0012 \pm 0.00023
\end{equation}
which is a five sigma discrepancy with theory.
It suggests a value $|\epsilon| \simeq 10^{-6}$
for the photon-mirror photon kinetic mixing\cite{fg}.
Taken at face value this experiment is strong evidence
for the existence of mirror matter and hence
parity symmetry. It is ironic that the last time
something important was discovered
in high energy physics with a table top experiment was
in 1957 where it was demonstrated that the ordinary
particles by themselves appear to violate parity symmetry.

Of course, this vacuum cavity experiment must be
carefully checked by another experiment to make
sure that mirror matter really exists. Actually
this is quite easy to do.  With the
largest cavity used in the experiment of Ref.\cite{vac}
the orthopositronium
typically collided with the cavity walls 3 times before
decaying. If the experiment was repeated with a larger cavity
then the mirror world effect would be larger
because the decohering effect of collisions would be
reduced. For example if a cavity 3 times larger could be used
(which means that the orthopositronium would typically collide
with the walls just once before decaying) then
the mirror world would predict an effect roughly 3 times larger.

There are several important implications of photon-mirror photon
kinetic mixing with the relatively large value of $|\epsilon| \simeq
10^{-6}$, some of which have been discussed
previously\cite{tunguska,fvpioneer,mph}. One very interesting effect is
that it allows mirror matter space-bodies to interact with the
Earth's atmosphere. Imagine that a mirror SB of velocity $v$ is
entering the Earth's atmosphere and plummeting towards the ground.
The mirror SB is constantly bombarded by the atmosphere in front
of it, initially with the velocity, $v$. Previous
work\cite{tunguska} has shown that the air molecules lose their
relative forward momentum after travelling only a distance of a
few centimeters within the mirror SB. The collision process is
dominated by Rutherford scattering of the 
atmospheric nuclei (of atomic number $Z \sim 7$) 
off the mirror SB nuclei (with mirror atomic number $Z'$)
of effective electric charge $\epsilon Z' e$.
The Feynman diagram for the scattering
process is shown in Figure 3. 
\vskip 0.4cm
\begin{center}
\epsfig{file=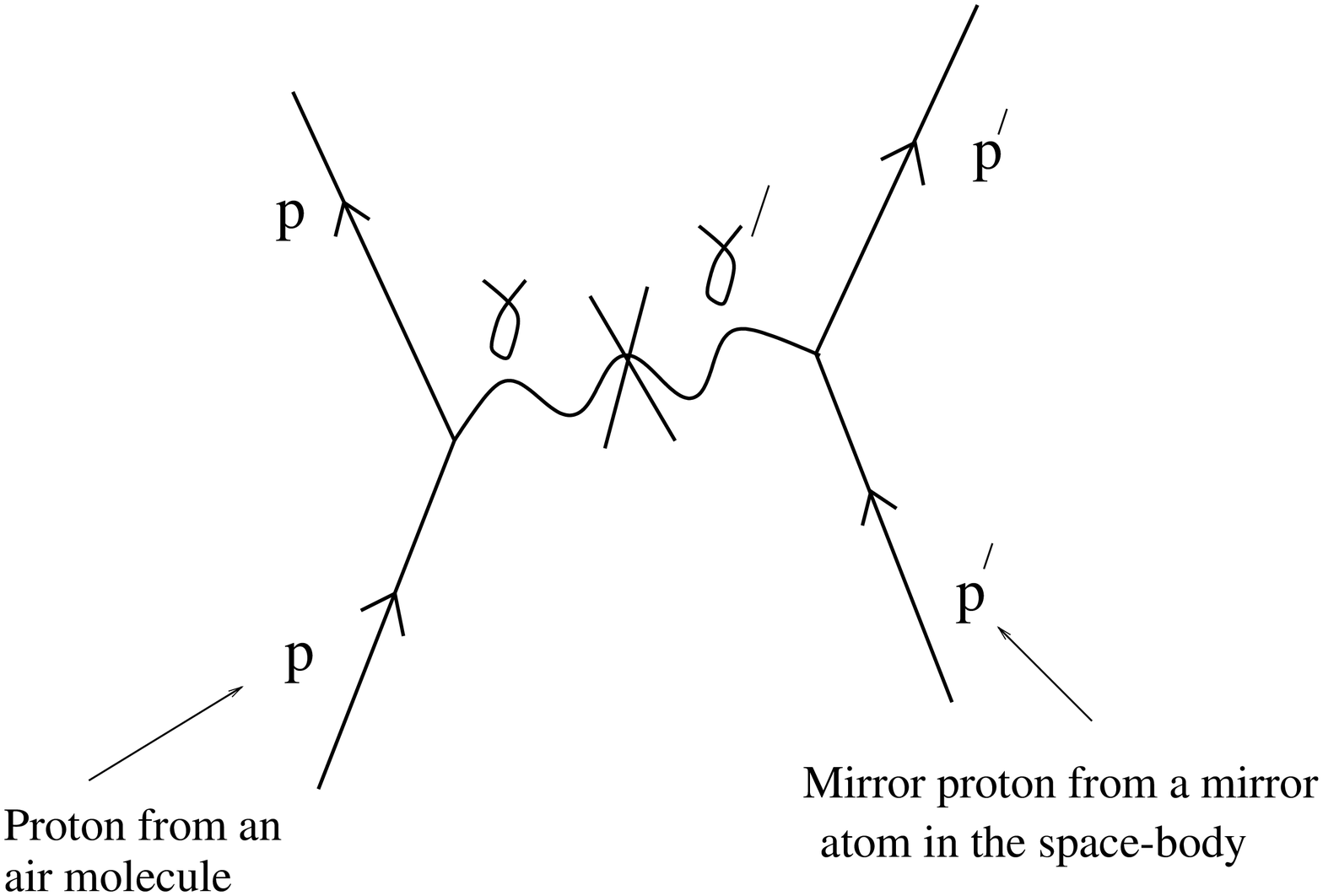,height=6cm,width=12cm}
\end{center}
\vskip 0.2cm
\noindent
FIG. 3. $\;$ Rutherford scattering of the mirror nuclei
off the atmospheric nuclei. The scattering is only possible because 
of the photon-mirror photon kinetic mixing, indicated by the cross (X)
in this diagram.
\vskip 0.8 cm 
\noindent
The interaction cross section is simply the standard Rutherford 
formula
(modified for small angle scattering by the screening effects of
the atomic electrons at the Bohr radius, $r_0 \approx 10^{-8}$ 
cm)\cite{mer} suppressed by a factor $\epsilon^2$:
\begin{eqnarray}\label{dsdOm}
    {d \sigma_{coll} \over d\Omega} = {4 M_A^2
    \epsilon^2 e^4 Z^2 {Z'}^2 \over (4M_{A}^2v^2
    \sin^2{\theta \over 2} + 1 /r_0^2)^2},
\end{eqnarray}
\vskip 0.4cm
\noindent
where $M_A$ is the mass of the air molecules\footnote{
Ordinarily, the Rutherford formula only applies
(for standard ordinary matter scattering) at high 
velocity ($v \stackrel{>}{\sim} 1000$ km/s) because
the Born approximation, from which it can be derived, is only
valid for weak potentials and high incident energies
(see e.g. Ref.\cite{mer}). 
In the case of ordinary-mirror matter scattering -- that
we are considering -- the potential is suppressed by
a factor of $\epsilon \sim 10^{-6}$, which means
that the Rutherford scattering formula is applicable
even for very low velocities such as $v \sim 1$ km/s.}.

Anyway, the Rutherford scattering causes the 
ordinary air molecules to lose their forward momentum within the
mirror space-body (assuming that $|\epsilon| \approx 10^{-6}$
as suggested by the experiments on orthopositronium\cite{fg}). 
It follows that the air resistance of a mirror SB is
roughly the same as an ordinary SB assuming the same trajectory,
velocity, mass and size and that the body remains intact. The
(kinetic) energy loss rate of the body through the atmosphere is then
\begin{eqnarray}\label{vx2}
&& {dE \over dx} =  {-C_d \rho_{atm}S v^2 \over 2},
\end{eqnarray}
where $\rho_{atm}$ is the mass density of the air, $v$ the speed, $S =
\pi R^2$ the cross sectional area and $R$ the effective radius 
of the mirror SB.
$C_d$ is an order of unity drag force coefficient depending on the
shape (and velocity) of the body. 
In Eq. (\ref{vx2}) the distance variable $x$ is
the distance travelled.

Equation (\ref{vx2}) is 
a standard result but we will derive it anyway.
Specifically, an on-coming air molecule which interacts with
the mirror SB and surrounding compressed air loses its relative
momentum, thereby slowing down the body. 
Conservation of momentum tell us that the change in
the SB velocity is then:
\begin{eqnarray}
\delta v = -{M_{A} \over M_{SB}} v.
\end{eqnarray}
Multiplying this by the number of air molecules [of
number density $n(h)$] encountered
after moving a distance $dx$, we have
\begin{eqnarray}
dv &=& -{M_A \over M_{SB}} v n(h) S dx
\nonumber \\
&=& - {\rho_{atm} v S dx \over M_{SB}}.
\label{ij2}
\end{eqnarray}
Note that this equation is equivalent to Eq.(\ref{vx2})
with $C_d = 2$. The factor $C_d$ arises because,
in general, not all air molecules in
the path of the body will lose their relative momentum
to the body; it is a complicated aerodynamic and hydrodynamic
problem, which depends on the shape, speed and trajectory of the body.
Solving Eq.(\ref{ij2}), we find an exponentially
decaying velocity:
\begin{eqnarray}
v = v_i e^{-x/D}
\end{eqnarray}
where $v_i$ is the initial velocity
of the SB and
\begin{eqnarray}
D = {x \over \int^x {\rho_{atm} S \over M_{SB}}dx}
\end{eqnarray}
For an air density of $\bar \rho_{air} \approx 10^{-3}\ g/cm^3$,
\begin{eqnarray}
D \sim 10 \left( {R \over 5 \ {\rm meters}}
\right)
\left( {\rho_{SB} \over 1 \ {\rm g/cm^3}}\right)
\ {\rm km.}
\label{good}
\end{eqnarray}

In general, one must also take into account the effect of mass loss
or `ablation'.
For an ordinary matter body, the air molecules
do not penetrate the body, but merely strike the surface
and bounce off. The energy is therefore dissipated right
at the surface which causes it to rapidly melt and vapourise.
This means that $R$ typically decreases quite rapidly for
an ordinary matter body.
For a mirror matter SB, some of the energy is dissipated within the body
by Rutherford scattering of the ordinary
air molecules with the
mirror atoms of the SB and also by collisions  of the air molecules
with other air molecules.
Furthermore the heating of the surrounding air as well
as the air trapped within the body should provide an efficient
means of transporting the heat. The air can transfer heat
from the surface regions of the mirror SB to the rest of the body.
As a crude approximation,
we can assume that the entire mirror matter SB, as well
as a significant fraction of air molecules within and surrounding
the SB are heated to a common temperature $T_b$.
We call this the `isothermal approximation'. 
Anyway, the important point is that in the mirror matter case, the
energy imparted to the SB is dissipated within it, rather than
just at its surface. Instead of rapid surface melting, the SB
initially has very low ablation (relative to the case
of an ordinary SB). The
kinetic energy of the impacting air molecules is dumped into the
mirror SB and surrounding air, and rapidly thermalized within it.

Broadly speaking two things can happen depending on the chemical
composition of the SB and also on its initial velocity and
trajectory: If the temperature of the mirror SB and surrounding
air reaches the melting point of the body, then the entire body
will break up and melt and subsequently vapourise. On the other
hand, if the temperature remains below the melting point, then the
body should remain intact. Note that once a SB breaks up into
small pieces it rapidly dumps its kinetic energy into the
atmosphere since its effective surface area rapidly increases,
which also rapidly increases the atmospheric drag force.

For a body that remains intact, there are two interesting
limiting cases. For large bodies with size much greater than 10
metres, the atmospheric drag force is not large enough to
significantly slow the body down during its atmospheric
flight, while for small bodies less than about 10 metres in size, they
typically lose most of their cosmic velocity in the atmosphere.
We will examine each of these limiting cases in turn,
starting with the most dramatic case of large bodies
such as the Tunguska SB.

\section{Heating of a large mirror space-body penetrating
the Earth's atmosphere}

In this section, we shall examine the case of a large mirror SB
entering the Earth's atmosphere. For large space-bodies, $R \gg
10$ metres, the atmospheric drag force does not slow them down
much [c.f. Eq.(\ref{good})], which means 
that we can treat their velocity as being
approximately constant during their atmospheric passage. 
For a SB
in an independent orbit around the Sun, the velocity at which the
body strikes the atmosphere (as seen from the Earth) is in 
the range\footnote{ The minimum
velocity of a space-body as viewed from Earth is not zero because
of the effect of the local gravity of the Earth. It turns out that
the minimum velocity of a space-body is about $11 \ $km/s, for a
body in an independent orbit around the Sun (and a little less if
there happened to be a body in orbit around the Earth).}:
\begin{eqnarray}
11 \ \hbox{km/s} \stackrel{<}{\sim} v \stackrel{<}{\sim} 70 \
\hbox{km/s.}
\end{eqnarray}
A pure mirror SB entering the Earth atmosphere would
have an extremely low initial temperature, only
a few degrees above absolute zero.
However, its temperature rapidly
begins to rise during its atmospheric passage as
the kinetic energy of the on-coming atmospheric
molecules (in the rest frame of the SB) are dumped into the
SB and the surrounding co-moving compressed air.
If the atmosphere was infinite in extent, the temperature
would eventually rise high enough so that the body
melts, in which case it would rapidly dump its
kinetic energy into the atmosphere because
the effective surface area of the body rapidly
increases when it breaks apart. The net effect would
be an atmospheric explosion. There is evidence that
such atmospheric explosions actually occur, and the
Tunguska event is one well studied example.

This mechanism has been discussed qualitatively
in Ref.\cite{tunguska}, and we wish now to examine
it quantitatively. Could it
reasonably be expected to occur
given what is known about the Tunguska event?
To answer this question we must first estimate the
rate at which energy is dumped into the mirror SB as it propagates
through the atmosphere.

We start with a simple model for the Earth's atmosphere.
We assume that it is composed of molecules of mass
$M_A \approx 30 M_p$ ($M_p$ is the proton mass),
with number density profile:
\begin{equation}\label{natm}
  n (h) = n_{0} \exp\bigg({-{h \over h_0}}\bigg),
\end{equation}
where $M_{A} n_0 \simeq 1.2 \times 10^{-3} $ g/cm$^3$ is the air
mass density at sea-level, and $h_0 \approx 8 \ $km is the scale
height. Eq.(\ref{natm}), which can be derived from hydrostatic
equilibrium, is approximately valid for $h \stackrel{<}{\sim}
25$ km. Above that height, the
density actually falls off more
rapidly then given by Eq.(\ref{natm}), but we will nevertheless
use this equation since it is a good enough approximation for the
things which we calculate.

The parameters defining the mirror SB's trajectory are illustrated
below in Figure 4. \vskip 0.7cm
\begin{center}
\epsfig{file=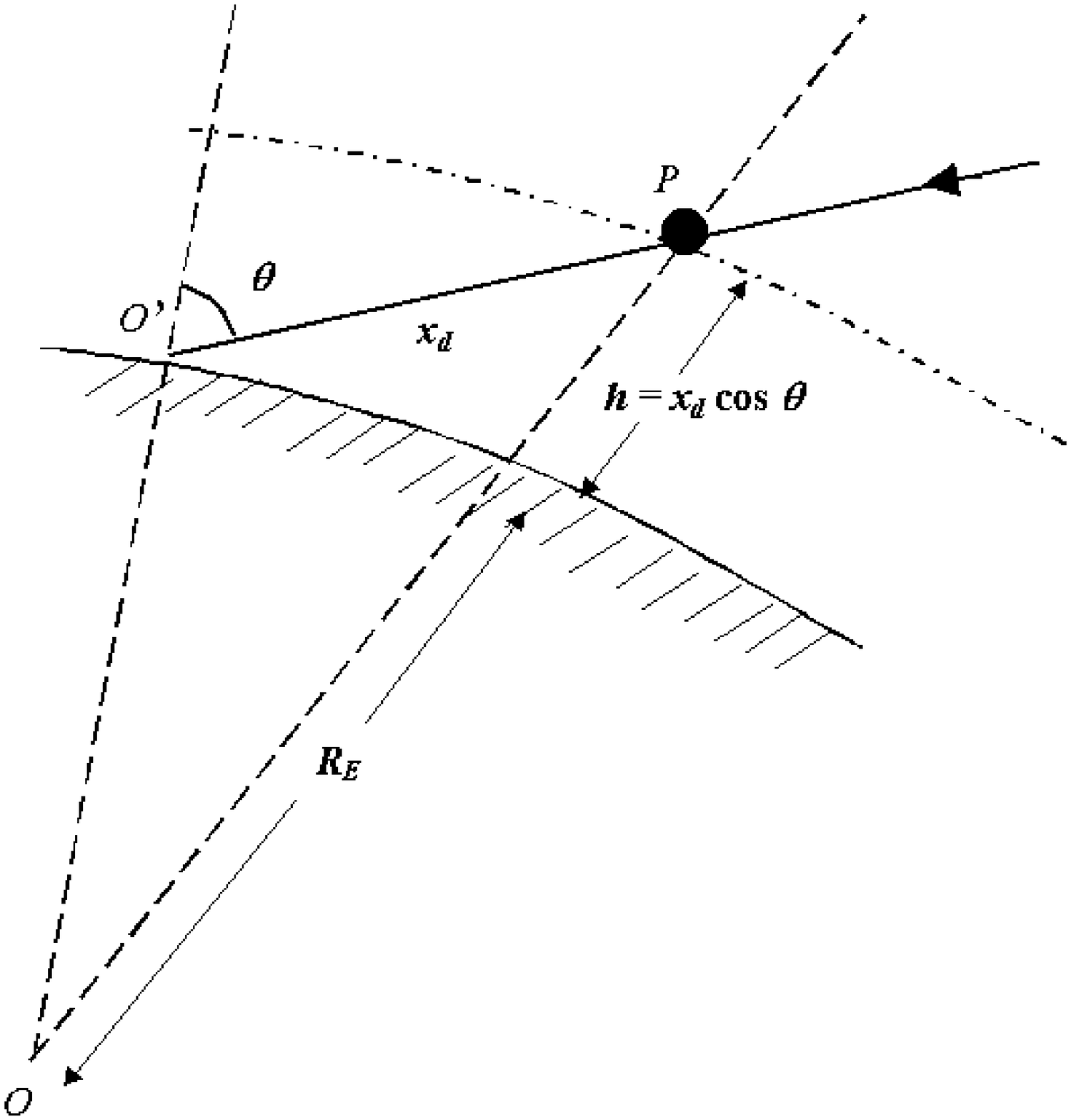,height=7cm,width=8cm}
\end{center}
FIG. 4. $\;$ Trajectory of a SB entering the Earth's
atmosphere, taken to be approximately a straight path. All the
length scales involved, $x, h$
%(and also $H$ which is not shown in the diagram)
are all very small compared to $R_E$,
allowing the curvature of the Earth be ignored. 

\vskip 0.8cm
\noindent
Its trajectory is directed towards a point on the ground
$O'$. Consider an instantaneous point $P$, located at a vertical height
$x_d \cos \theta$ above the ground, where the distance $O'P$ is $x_d$.
In the frame of the SB, on-coming air molecules strike the body
and/or surrounding compressed air (both outside and within the mirror body),
eventually losing most of their kinetic energy ($M_A v^2/2$) after many
collisions.  Their kinetic energy is converted primarily into 
thermal energy, heating the body and surrounding compressed air.  

To estimate
the amount of energy dumped into the mirror SB
and surrounding compressed air
from the interaction of it with the atmosphere, consider
an infinitesimal distance $dx$
(at the point $P$).
As a simple approximation, let us assume that all
the air molecules in the volume $S dx$ are swept
up by the on-coming SB and surrounding air.
This approximation is similar to the one
leading to Eq.(\ref{ij2}) 
where detailed hydrodynamic and aerodynamic effects
are ignored (equivalent to setting $C_d = 2$).
In this approximation,
the energy, $d\epsilon$, dumped into the mirror SB and 
surrounding compressed air is simply given by
the number of air molecules in the volume $Sdx$ multiplied
by their average energy (with respect to the rest frame
of the SB):
\begin{equation}\label{depsilon}
d\epsilon = n(h) {1 \over 2} M_A v^2 S dx .
\end{equation}
This means that the total energy dumped into the space-body (and
surrounding highly compressed co-moving air) during its passage
from far away up until the point $P$ is simply:
\begin{equation}
\epsilon (x_d) = \int^{\infty}_{x_d} n_0 exp\left({-x\cos\theta
\over h_0}\right) {1 \over 2} M_A v^2 S dx . \label{y1}
\end{equation}

We are most interested in working out the energy going into
just the mirror SB, rather than both SB and surrounding compressed
air.  Actually this 
is another difficult hydrodynamic problem.
The air trapped within the body and in front of
it is highly compressed. The total number
of air molecules moving with the SB could be of the same
order as the total number of mirror molecules within the SB.
With our isothermal assumption, we can lump this 
hydrodynamic uncertainty into a single factor, $f_a$, which is just
the fraction of SB molecules to air molecules co-moving with
the SB ($N_{air}$):
\begin{eqnarray}
f_a =
{N_{SB} \over N_{SB} + N_{air}}.
\end{eqnarray}
Here $N_{SB}$ is the number of mirror molecules comprising the SB.
If these molecules have mass $M_{A'}$ (for example,
$M_{A'} \simeq 18M_P$ for mirror H$_2$O ice) 
then $N_{SB} = M_{SB}/M_{A'}$.
Essentially $f_a$ is the the proportion of the kinetic energy
of the on-coming air molecules transferred into heating the mirror SB.
The energy dumped into the SB is then
\begin{eqnarray}
\epsilon_{SB} (x_d) &=& f_a\epsilon (x_d) 
%\nonumber \\ &=&
%\int^{\infty}_{x_d} f_a n_0 \exp\left({-x\cos\theta \over
%h_0}\right) {1 \over 2} M_A v^2 S dx.
\end{eqnarray}

In addition to the factor $f_a$, there are other hydrodynamic uncertainties
coming from the flow of air around the body. We assumed
that every air molecule in the path of the body would
be swept up by the body, however real life is always more
complicated. 
In general we must model the flow of air around the body --
a difficult hydrodynamical problem... So, we must introduce
another hydrodynamic uncertainty, $f_b$. Actually, we will
combine all of our hydrodynamic
uncertainties ($f_a$ and $f_b$) into a single factor 
$f$.
We will later take $f \sim 0.1$, but this is quite uncertain.
A useful quantity is the energy gained per molecule of the
space-body, $\stackrel{\sim}{\epsilon}_{SB} = \epsilon_{SB}
(x_d)/N_{SB}$. Evaluating this quantity, we find:
\begin{eqnarray}
\stackrel{\sim}{\epsilon}_{SB} &=& {3fn_0 v^2 M_A M_{A'} h_0
\exp\left({-x_d \cos \theta \over h_0}\right) \over 8 \rho_{SB} R
\cos \theta}, \nonumber
\\
&\simeq & {5 f \over \cos\theta}\left( {M_{A'} \over 18M_P}\right)
\left( {100 \hbox{ m} \over R}\right) \left( {1 \hbox{g/cm}^{3}
\over \rho_{SB}}\right) \left( {v \over 30 \ \hbox{km/s}}\right)^2
\exp\left({-x_d \cos \theta \over h_0}\right) \ \hbox{eV,}
\end{eqnarray}
where $M_P$ is the proton mass and $R$ is the effective radius
of the mirror SB.

If we know
the energy dumped into the SB, $\epsilon_{SB} (x_d)$,
we can estimate the temperature gained by the SB if we know
the specific heat of the body. Recall that specific heat is
just the energy required to raise the temperature of a
body by 1 degree. Actually we are particularly interested
in the energy required to heat the body from a temperature
near absolute zero until it melts. For this we must
integrate the specific heat from near absolute zero to
the melting temperature, and also add on the heat of fusion,
which is the energy required to effect the phase transition.
This total energy -- which we label $\epsilon_m$ -- obviously
depends sensitively on the type of material.
While we do not have
any direct empirical guidance about the likely chemical compositions
of mirror SB, we can be guided by the compositions of
ordinary matter SB -- the asteroids and comets.
An important observation though is that while icy ordinary
matter bodies (comets) have only a relatively short lifespan in the
inner solar system because they would have been melted
by the Sun (and therefore can only exist in
elliptical orbits), mirror space-bodies made of 
mirror ices could be plentiful in the 
inner solar system (in circular orbits) and might be expected to dominate 
over non-volatile substances.
In the table
below, we estimate this quantity for a few plausible
space-body materials.

%%%%%%%%%%%%%%%%%%%%%%%%%%% Table %%%%%%%%%%%%%%%%%%%%%%%%%%%%%%%%%

\vskip 1cm
\begin{center}
{\bf Table I:} \ Physical properties \cite{crchb} of some cited
minerals in common meteorites\cite{table}
\end{center}

\vskip 0.5cm 
 
\begin{center}
\begin{tabular}{|c|c|c|c|c|c|}  \hline
\setlength{\tabcolsep}{1pt} \label{tabella1} Mineral ($A'$) &
$\rho_{SB}$ & $T_m$ & $Q_1$ & $L_F$ & $\epsilon_m$
\\   & (g/cm$^3$) & (K)  &  (kJ/mol) & (kJ/mol) & (kJ/kg) [eV per molecule] \\
\hline
\hline
Ammonia Ice, NH$_3$ (17) & 0.8 &195 & 4.2 &5.7 &580 (0.1) \\
Methane Ice, CH$_4$ (16) & 0.5 &91 & 2.6 &0.9 &220 (0.04)
\\
Ice, H$_2$O
(18) &0.9 &273 &6 &6 &670 (0.12) \\
Cristobalite, SiO$_2$ (60) &2.2 &1996 &120 &9.6 & 2200 (1.4) \\
Enstatite, MgSiO$_3$ (104) &3.9 &1850 & 200& 75 $\pm$ 21 & 2500
(2.8) \\
Forsterite, Mg$_2$SiO$_4$ (140) &3.2 &2171 & 360 &71 $\pm$ 21 &
3000 (4.5) \\
Fe (56) &7.9 &1809 &62 &13.8 &1400 (0.8) \\
Magnetite, Fe$_3$O$_4$ (232) &5.2 &1870 &360 &138 & 2100 (4.1) \\
Troilite, FeS (88) &4.7 &1463 &88 &31 &1400 (1.2) \\
Nickel (59) &8.9 &1728 &53 &17 &1200 (0.7) \\ \hline
\end{tabular}
\end{center}
\vskip 0.3cm
\noindent
$A'$ denotes mirror atom's relative molecular weight; \\ $T_m$
denotes melting point of $A'$; \\ $Q_1$ denotes total heat
absorbed by $A'$ to raise its temperature from 0 $^o$K to its melting
point; \\ $L_F$ denotes heat of fusion (from solid phase to liquid
phase);
\\ $\epsilon_m = Q_1 + L_F$.
\vskip 1 cm
%%%%%%%%%%%%%%%%%%%%%%%%%%%%%%%%%%%%%%%%%%%%%%%%%%%%%%%%%%%%5

Assuming that no fragmentation occurs (prior to melting), then a
large mirror SB would melt
above the ground provided that $\stackrel{\sim}{\epsilon}_{SB}(x_d = 0)
> \epsilon_m$.
Solving this condition for the SB velocity, $v$, we find:
\begin{eqnarray}\label{largevini}
  v \stackrel{>}{\sim}&& 10 \sqrt{\left({\cos \theta \over 0.5}\right)
\left( {0.1 \over f}\right)
%\nonumber \\ &&
  \bigg({R \over 100 \hbox{ m}}
  \bigg)
  \bigg({18M_P \over M_{A'}}\bigg)
    \bigg({\rho_{SB}\over 1 \hbox{ g/cm}^3}\bigg)
\bigg( {\epsilon_m \over 0.1 \ \hbox{eV}}\bigg) }
    \ \hbox{ km/s}.
\end{eqnarray}
We emphasise that this equation was derived assuming that the
body's velocity, $v$, is approximately constant. This is roughly
the case for a large body ($R \gg 10$ metres) because it is not
slowed down much by the atmospheric drag force [c.f. Eq.(\ref{good})]. 
Observe that
Eq.(\ref{largevini}) suggests that a large ($R \sim 100 $ m)
mirror SB made of mirror ices would typically melt at some point
in the atmosphere (only for $R \gg 100$ m could a mirror icy body
survive to hit the ground). Once it melts, the effect of the
pressure of the atmosphere on the liquid body would cause it to
disperse dramatically, increasing its effective surface area. This
greatly increases the atmospheric drag force, causing the body to
rapidly release its kinetic energy into the atmosphere, leading in
essence, to an atmospheric explosion.

On the other hand, a large mirror rocky body could 
survive to hit the ground intact; only if its velocity was
relatively large ($\stackrel{>}{\sim} 50$ km/s) would it 
melt in the atmosphere. Of course, our estimation might
be too simplistic to allow us to draw very rigorous
conclusions. Nevertheless, the mirror matter hypothesis does seem
to provide a nice explanation for the main features of the
Tunguska event: the low altitude explosion, absence of ordinary 
fragments, and no chemical traces. Let us now take a closer look.

The SB would melt at a height $h = x_d \cos\theta$
above the ground if
$\stackrel{\sim}{\epsilon}_{SB}(x_d) =
\epsilon_m$.
Solving this condition for the height, $h$, we find
\begin{eqnarray}
h = h_0 \ln \left[ \left( {v \over 10 \ \hbox{km/s}} \right)^2
\left( {1 \hbox{g/cm}^3 \over \rho_{SB}}\right) \left( {100 \hbox{
m} \over R}\right)\left( {f \over 0.1}\right)\left({0.5 \over
\cos\theta }\right) \left( {M_{A'}\over 18M_P}\right)\left( { 0.1 \
\hbox{eV} \over \epsilon_m}\right) \right].  \label{12}
\end{eqnarray}
Focussing our attention onto the Tunguska event, which is characterized by
$M_{SB} \sim 10^8 - 10^{9}$ kg ($\Rightarrow R \sim 40$ m for mirror ice 
and 20 m for mirror iron), $\ \cos\theta \sim 0.5$, and the height of
the atmospheric explosion is $h \sim h_0$. In this case we find
that
\begin{eqnarray}
v &\sim & 12 \ \hbox{km/s} \ \ {\rm for \ mirror \ ice,}\nonumber
\\ v &\sim & 40 \ \hbox{km/s} \ \ {\rm for \ mirror \ iron.} \label{13}
\end{eqnarray}
Recall the range of expected velocities of the SB
is $11 \hbox{ km/s} < v < 70
\hbox{ km/s}$.
Thus, it seems that both mirror ices and mirror non-volatile material
may plausibly explain the Tunguska event.

An interesting observation is that if large mirror SB are 
predominately made of mirror ices, then Eq.(\ref{12}) suggests that
smaller such bodies should explode at higher altitudes because of
their smaller $R$ values. Roughly speaking, the energy gained per
molecule (in a large SB) is proportional to the area swept out divided by the
number of SB molecules (i.e. $\propto S/V \propto 1/R$). That is, the
energy gained per molecule is inverserly proportional to the
body's size. Thus, smaller bodies should heat up faster.
Furthermore, the energy gained also depends on the body's
velocity, but the characteristics of the Tunguska event suggest a
mirror body near the minimum value possible, $11$ km/s [see
Eq.(\ref{13})]. SB with higher velocities, which are possible,
would also lead to higher altitude explosions.
Thus, the unique low altitude explosion associated with the Tunguska
event seems to have a simple explanation in this
mirror matter interpretation.
Smaller mirror SB of Tunguska chemical composition should
always melt and thereby explode higher up in the atmosphere.
This feature is in accordance with the observations
of airburst events discussed in part II.

To conclude this section, let us mention that there
are many puzzling features of the Tunguska event which
we have yet to address here.
For example, the origin of the optical anomalies, including
abnormal sky-glows and unprecedented bright noctilucent clouds
\footnote{
These optical anomalies are in addition to the visual sightings 
of the bolide on June 30, 1908. The visual sightings
could be explained in the mirror SB
hypothesis because part of the kinetic energy of the SB is 
transferred to the air, which is eventually converted
into ordinary heat and light as the high velocity air molecules 
(at least 11 km/s) eventually interact with (more distant) 
surrounding `stationary' air.
}.
The most puzzling aspect of which is that they seem to appear
a few days {\it before} the Tunguska event (for a review,
see e.g. \cite{vasilyev}).
We make no claims that the mirror
matter space-body hypothesis definitely explains all of the observed
effects; our main task is to see if it can explain
the main characteristics of the event (the low altitude 
atmospheric explosion, absence of fragments and chemical traces, 
visual sightings of the bolide). 
We suggest that the broad features of a SB made of mirror
matter are indeed consistent with the main features of the 
Tunguska event. Further work needs to be done to see
to what extent it can explain 
the other reported features.

\section{Heating of a small mirror space-body penetrating
the Earth's atmosphere}

In this section we will examine the case of a small mirror SB
entering the Earth's atmosphere. For small bodies, $R
\stackrel{<}{\sim} 10$ metres, the atmospheric drag force
effectively ``stops'' the body in the atmosphere -- it loses its
cosmic velocity and would reach the Earth's surface with only a
relatively low impact velocity in the range $0.1 \ \hbox{km/s}
\stackrel{<}{\sim} v_{impact} \stackrel{<}{\sim} 3 \ \hbox{km/s}$
(providing of course that it doesn't melt or break up on the way
down). 

Because the velocity is not constant in this case,
we must combine Eq.(\ref{depsilon}) with Eq.(\ref{ij2}):
\begin{eqnarray}
d\epsilon &=& n(h) {1 \over 2} M_A v^2 S dx \nonumber \\
dv &=& -{M_A \over M_{SB}}v n(h) S dx.
\end{eqnarray}
That is,
\begin{eqnarray}
d\epsilon = -{v M_{SB} \over 2} dv.
\end{eqnarray}
Thus, if the body loses most of its cosmic velocity in
the atmosphere, then integrating the above equation
(and putting in the hydrodynamic factor, $f$), we find:
\begin{equation}
\epsilon_{SB} \approx f M_{SB} v^2/4.
\end{equation}
Here, $v$ is the initial velocity of the SB.

The above equation for the energy transferred to heating the SB
can also be conveniently expressed in terms of heat energy per
molecule, $\stackrel{\sim}{\epsilon}_{SB} \equiv
\epsilon_{SB}/N_{SB}$:
\begin{eqnarray}
\stackrel{\sim}{\epsilon}_{SB} &=&  f M_{A'} v^2/4 \nonumber \\
&\simeq & f \left( {M_{A'} \over 18 M_P}\right) \left( {v \over 11 \
\hbox{km/s}}\right)^2 5 \ \hbox{eV.} \label{mon}
\end{eqnarray}
Clearly, the only small mirror space-bodies which can survive to
strike the ground without completely melting must have relatively
high values for $\epsilon_m$ and relatively low initial velocities
near the minimum $\sim 11$ km/s. 
In the following table we compare $\stackrel{\sim}{\epsilon}_{SB}$
with $\epsilon_m$ for some plausible mirror SB materials.
\vskip 1.0cm
%%%%%%%%%%%%%%%%%%%%%%%%%%%%%%%% Table 2 %%%%%%%%%%%%%%%%%%%%%%%%%%%
\begin{center}
{\bf Table II} \, Comparison between $\epsilon_m$ and $\tilde
\epsilon_{SB}$ for a small mirror SB ($R \stackrel{<}{\sim} 10$
metres).
\end{center}
\vskip 0.2cm
\begin{center}
\begin{tabular}{|c|c|c|}  \hline
\setlength{\tabcolsep}{1pt} \label{tabel2}
   Mineral & $\tilde \epsilon_{SB}$ (eV per molecule)& 
$\epsilon_m$ (eV per molecule)
\\ \hline \hline H$_2$O \& NH$_3$ Ice &
$0.5\left( {f \over 0.1}\right)\left({v \over 11 km/s}\right)^2 $
&0.1
\\ CH$_4$ Ice &
%(H$_2$0, CH$_4$ and NH$_3$) &
$0.5\left( {f \over 0.1}\right)\left({v \over 11 km/s}\right)^2 $
&0.04
\\
Cristobalite, SiO$_2$ & $1.5\left({f \over 0.1}\right)\left({v \over
11 km/s} \right)^2   $ & 1.4
\\
Enstatite, MgSiO$_3$ & $2.5\left({f \over 0.1}\right)\left({v \over
11 km/s} \right)^2   $
 &2.8 \\
Forsterite, Mg$_2$SiO$_4$ & $3.5\left({f \over 0.1}\right)\left({v
\over 11 km/s} \right)^2   $ &4.5  \\
Fe & $1.5\left({f \over 0.1}\right)\left({v \over 11 km/s} \right)^2
$ &0.8
\\
Magnetite, Fe$_3$O$_4$ & $7\left({f \over 0.1}\right)\left({v
\over 11 km/s} \right)^2   $ &4.1
\\
Troilite, FeS & $2\left({f \over 0.1}\right)\left({v \over 11
km/s} \right)^2   $ &1.2\\
Nickel & $1.5\left({f \over 0.1}\right)\left({v \over 11 km/s}
\right)^2   $ &0.7  \\
 \hline
\end{tabular}
\end{center}
%%%%%%%%%%%%%%%%%%%%%%%%%%%%% end table 2 %%%%%%%%%%%%%%%%%%%%%%
\vskip 0.7cm

\noindent As this table shows, a small mirror SB can potentially
reach the Earth's surface without melting provided that $f
\stackrel{<}{\sim} 0.1$ if they are made of typical non-volatile
materials (and $f \stackrel{<}{\sim} 0.02$ if the are made of
ices). While values of $f$ as low as $\sim 0.1$ are presumably
possible
\footnote{ Roughly speaking Eq.(\ref{mon}) also applies to
ordinary matter bodies which lose their cosmic velocity in the
atmosphere; parts of which certainly do sometimes survive without
completely melting. In addition to meteorite falls, there are also
well documented space debris, such as parts of satellite (e.g.
Skylab debris). Satellite parts enter the Earth's atmosphere at
about $8$ km/s, just below the minimum velocity of SB's.}, 
values
as low as $0.02$ seem less likely. This means that a small
mirror matter SB can possibly survive to hit the Earth's surface
provided that it is made of a non-volatile composition such as mirror
rocky silicate materials or mirror iron materials.

Can such a small mirror matter SB be responsible for the observed
anomalous events discussed in section IIB?
Recall that 
an ordinary matter explanation for these events suffers from two
main difficulties.
First, any low altitude ordinary
body must have had a large highly luminous parent body. Second,
a body having survived to low altitudes, losing its cosmic velocity, 
should have left recoverable fragments. Remarkably both of these
difficulties evaporate in the mirror matter interpretation.

First, the lack of ordinary fragments is easily explained
if the SB is made of mirror matter. One might
expect recoverable mirror matter fragments, but
these might have escaped notice (especially if $\epsilon$
is negative, as we will explain in the next section).

Second, low altitude mirror SB need not have been large and highly
luminous at high altitudes. Ablation should occur at a much lower
rate for a mirror SB because the pressure of the atmosphere is
dissipated within the body (rather than just at its surface as
would be the case for an ordinary matter body entering the
atmosphere). 
More importantly,
air can be trapped {\it within} the body and can
transport the heat away from the surface regions.
The mechanism for producing ordinary light from a
mirror SB is therefore completely different to an ordinary matter
body. For a mirror SB, we can identify three basic 
mechanisms for producing ordinary light:
\vskip 0.2cm
\begin{itemize}
\item
Interactions with the air through ionizing collisions
(where electrons are removed from the atoms) and excitations
of the air molecules.
\item
The potential build up of ordinary electric charge as a
consequence of these ionizing collisions which can trap ionized
air molecules within the body. This build up of charge can lead to
electrical discharges\cite{ceplecha}. Note that this trapping of
air cannot
occur if the SB is made of ordinary matter.
\item
Heating of any ordinary matter fragments (if they exist!)
within the mirror
matter space-body, which subsequently radiates ordinary light.

\end{itemize}

The first two of these mechanisms listed above are actually most
important for very speedy (mirror) meteoroids (as we will explain in
a moment) but nevertheless may still play a role even if the
velocity is near the minimum $\sim 11$ km/s as we suspect to be
the case for the anomalous low altitude fireballs (such as the
2001 Jordan event and the 1994 Spanish event). A small mirror matter
body can thereby be relatively dim at high altitudes (especially
if it is of low velocity, $v \sim 11$ km/s). As it moves through
the atmosphere it slows down due to the atmospheric drag force.
The kinetic energy of the body is converted into the heating of
the whole body and the surrounding compressed air both outside and
{\it within} the body. Any ordinary matter fragments within the
body will also heat up and emit ordinary light, but the body need
not be extremely bright at high altitudes. The heating of the
whole body could make the body act like a heat reservoir, perhaps
allowing 
trapped air and ordinary matter fragments
to emit some ordinary light even at low
altitudes where it is moving relatively slowly
(especially as our estimates in table II suggest that the
body will be heated to near melting temperature which is
typically $\stackrel{>}{\sim} 1800\ K$  for plausible non-volatile
mirror SB materials).

Let us briefly expand upon the effect of ionizing collisions and
the potential build up of ordinary charge within the SB due to
trapped ionized air molecules. This mechanism is most important at
very high altitudes ($\stackrel{>}{\sim} 100 \ $km), since impacting air molecules
can strike the mirror SB with their full velocity. At lower
altitudes compressed air develops within and in front of the body
which can shield the mirror SB from direct impacts at cosmic
velocity. Note that this mechanism is also most important for SB
entering the atmosphere with high velocity. The impacting air
atoms have energy,
\begin{eqnarray}
E = {1 \over 2}M_{A}v^2 \approx 70 \left({v \over 30 \hbox{
km/s}}\right)^2 \ \hbox{eV}.
\end{eqnarray}
For a high velocity SB, $v \sim 70$ km/s, the energy of the impacting air
atoms is sufficiently high for ionizing collisions to
occur. For ordinary SB, the probability of ionizing collisions at
these velocities is
quite low\cite{86partdatabook}, however for mirror SB energy loss
due to ionizing collisions can be potentially 
comparable to Rutherford scattering. This might
explain the anomalously high altitude
beginnings of the observed light from certain speedy
meteors\cite{ceplecha}. 
It might also explain the release of
electrical energy in some anomalous events such as the 1993 Polish
event briefly mentioned in section IIB, because ionizing
collisions should lead to a build-up of ordinary electric
charge within the mirror SB from trapped ionized air
molecules.

Of course, if the body is made (predominately) of mirror matter and
does survive to hit the ground,
it would not leave any significant ordinary
matter fragments. This obviously simply explains the other
mysterious feature of the
anomalous small fireball events -- the lack of ordinary fragments
(despite the fact that the body was actually observed
to hit the ground at low velocities).
An important consequence of the mirror matter interpretation of
the anomalous small fireball events is that
mirror matter should exist in solid form (if they
are indeed due to non-volatile mirror matter material\footnote{
Transfer of heat from the air to the mirror body would
quickly melt any volatile mirror matter components (if they are on
the Earth's surface), even 
if they could survive to hit the ground.}) and may therefore
be potentially recoverable from these impact sites as
we will explain in the following section.

\section{Finding mirror matter in/on the ground and the sign of 
$\epsilon$}

The photon-mirror photon kinetic
mixing induces small ordinary electric charges
for the mirror electron and mirror proton.
A very important issue though is the effective sign
of this induced ordinary electric charge (the
orthopositronium experiments are only sensitive
to $|\epsilon |$, they don't provide any information
on the sign of $\epsilon$).
There are basically two physically distinct possibilities:
The induced charge is either of the same sign or
opposite, that is,
either the mirror electrons repel ordinary electrons,
or they attract them.
In the case where ordinary and mirror electrons have
ordinary charge of the same sign, the ordinary and
mirror matter would repel each other.
%\footnote{ This assumes that we have mirror matter
%not mirror anti-matter. If we have mirror anti-matter
%rather than matter, than things are reversed. That is, xxxxx}.
In this case,
a fragment of mirror matter could remain
on the Earth's surface, largely unmixed with ordinary matter.
In the case where the mirror electrons have a tiny ordinary
electric charge of the opposite sign to the ordinary electrons,
the mirror atoms attract the ordinary ones.
In this case it would be energetically favourable for
mirror matter to be completely immersed in ordinary matter,
releasing energy in the process.

In the first case, the maximum repulsive force at the
mirror matter - ordinary matter boundary can
be crudely estimated to be of order\footnote{
This equation completely neglects the shielding effect of
electrons which means it is probably an over estimate (by
an order of magnitude or two) of the maximum electrostatic
repulsive force. Nevertheless, it is perhaps good enough
for our purposes.}:
\begin{eqnarray}
F^{electrostatic}_{maximum}
&\sim & N_{atoms}^{surface} {\epsilon Z_1 Z_2 e^2 \over r_{bohr}^2}
\nonumber \\
&\sim & \left( N_{SB}\right)^{2/3} {\epsilon Z_1 Z_2 e^2 \over r_{bohr}^2},
\end{eqnarray}
where $N_{atoms}^{surface}$ is the number of surface atoms
[which is related to the total number of atoms, $N_{SB}$ by,
$N_{atoms}^{surface} \sim \left(N_{SB}\right)^{2/3}$]
and $Z_1$ ($Z_2$) is the atomic number of the
ordinary (mirror) nuclei.
The electrostatic force opposes the force of
gravity, so a mirror rock can be supported on
the Earth's surface provided that
$F^{electrostatic}_{maximum} > F^{gravity}$.
$F^{gravity}$ is simply given by
\begin{eqnarray}
F^{gravity} &=&
g M_{rock} \nonumber \\
&=& g M_{A'} N_{SB},
\end{eqnarray}
where $g \simeq 9.8 \ {\rm m/s^2}$ is the acceleration of
gravity at the Earth's surface.
Thus, we find:
\begin{eqnarray}
{F^{gravity} \over F^{electrostatic}_{maximum}}
\sim
{g M_{A'} r_{bohr}^2\left(N_{SB}\right)^{1/3}
\over
\epsilon Z_1 Z_2 e^2 }.
\end{eqnarray}
For a macroscopic sized body, $N_{SB}$ is of order
the Avagadro's number, $N_A$. Putting in the numbers we find:
\begin{eqnarray}
{F^{gravity} \over F^{electrostatic}_{maximum} }
\sim
10^{-6}\left({10^{-6} \over \epsilon }\right) \left({N_{SB} \over N_{A}
}\right)^{1/3},
\end{eqnarray}
where we have taken $Z_1 \sim Z_2 \sim 10$ and $M_{A'} \sim 20M_P$.
Clearly, in this case where the induced ordinary electric charge
of mirror particles are of the same sign as their ordinary
counterparts, mirror matter bodies can be supported against
gravity if $\epsilon \sim 10^{-6}$ as suggested by experiments
on orthopositronium\cite{fg}. If they impact with 
the Earth at low velocity then one
might expect mirror fragments to exist right on the ground at the
impact sites (or perhaps partly embedded in the ground). 

Of course,
a pure mirror rock or fragment would be invisible, but
one could still touch it and pick it up (if $\epsilon \sim 10^{-6}$).
If it contained embedded ordinary matter, then it would
then be visible but surely of unusual appearance.
Of course, the fact that no such mirror rocks have been
found may be due to their scarcity.
The oldest terrestrial age of an iron meteorite that has been
recovered on Earth (in Tamarugal in the Atacama desert, Chile) is
1.5 million years\cite{yoonbook}. This represents a crude upper 
estimate for the typical time that an extraterrestrial body could 
remain and be found on the Earth's surface
before it is buried by
tectonic movements of the Earth's crust. The actual
rate of anomalous low altitude bolide events across the entire
Earth should be roughly 10-100 times the observed event rate 
(since they are only observed in populated regions).
Thus, since Jordan-like events are observed to occur roughly on a yearly
basis\footnote{
In this estimate, we 
use the smaller Jordan-type events discussed in section IIB, rather
than large Tunguska-type atmospheric explosions
(section IIA), because the former seem to be more likely
to leave large non-volatile fragments (`miror rocks') 
which should be easier to find then small fragments which
might be left after atmospheric explosions. 
}, one might expect them to occur on the planet roughly
once every few weeks. Each event might leave small rocks
and fragments over an area of about 1 m$^2$. 
Thus, the total area covered by mirror rocks and fragments 
integrated over the 
last million years is roughly expected to be 
less than of order
$10^{7}-10^{8}$ m$^2$ (or
between 10-100 square kilometers) -- a very small fraction of the Earth's
surface. Nevertheless, the odds of finding a mirror matter rock could
be greatly improved if one were to carefully search the `impact
sites', such as the one in Jordan and the one in Spain.
Nevertheless, the fact that no strange invisible rocks have
been recovered suggests perhaps that 
the induced ordinary electric charge of mirror
particles is of {\it opposite} sign as their ordinary
counterparts. In this case things are more interesting but
somewhat more complicated.

If ordinary and mirror matter attracts each other, as it would do
if say $\epsilon $ is negative (or if one had a piece of mirror
anti-matter and $\epsilon$ were positive) then a fragment of
mirror matter should become completely immersed in the ground (or
at least the bulk of it). In the process energy will be released,
but exactly how much is a non-trivial solid state
problem.
%\footnote{ A very amusing possibility is the case where
%part of the mirror body protrudes from the Earth (and $\epsilon < 0$). In 
%this case, we could embed an ordinary sword in it. In the process energy would
%be released (most in the form of heat). However, energy would be
%required to remove the sword from the rock. Thus, the legendary
%``sword in the stone'', may actually be possible if the stone is
%made of mirror matter, nevertheless, this legend is probably still
%fictitious...}. 
In the realistic case of materials of varying
density and composition, a mirror body should be expected to come
to rest within the Earth, probably only just below the surface
($\sim$ meters).

Thus, irrespective of the sign of $\epsilon$, mirror matter should be
found in the impact sites -- either on the ground or buried beneath the
surface -- if the anomalous low altitude fire balls are indeed
caused by mirror SB. In some cases (e.g. the 2001 Jordan event)
these sites are highly localised and easily accessible. One could
collect samples of earth underneath the impact site and try to search
for the presence of mirror matter fragments in the sample.
Although chemically inert, mirror matter still has mass
so its presence can be inferred by its weight. If one
could count the number of ordinary atoms (and their mass) in the sample,
maybe by using a mass spectrometer, then the presence
of invisible gravitating matter could be inferred.

%%% additional stuff here %%%

Another complementary way to test for mirror matter is by
searching for its thermal effects on the ordinary matter surroundings.
Consider the case where there is indeed mirror matter embedded in
the ground. The mirror matter will be heated up by the interactions
with the ordinary atoms. However,
the mirror matter will thermally radiate mirror photons (which quickly
escape) providing a cooling mechanism.
Heat will be replaced from the ordinary matter surroundings which will
become cooler as a result.

To make this quantitative, consider a spherical mirror
body of radius $R'$ surrounded by a much larger volume
of ordinary matter. In this case, the energy lost per unit time to
mirror radiation is given by the Stefan-Boltzmann law:
\begin{eqnarray}
Q_{rad} = \sigma T'^4 A',
\end{eqnarray}
where $T'$ is the surface temperature of the mirror body and $A'$ 
is its surface area.
As energy escapes, this causes a temperature gradient in
the surrounding ordinary matter as it thermally conducts heat
to replace the energy lost.
If we consider a shell of radius $R$ surrounding
the mirror body ($R > R'$) then
the heat crossing this surface per unit time is given by:
\begin{eqnarray}
Q = -\kappa {\partial T \over \partial R} A,
\end{eqnarray}
where $\kappa$ is the coefficient of thermal conductivity,
and $A = 4 \pi R^2$.
Thus, equating the thermal energy transported to
replace the energy escaping (as mirror radiation), we 
find:
\begin{eqnarray}
{\partial T \over \partial R} =
{-\sigma T'^4 R'^2 \over \kappa R^2}.
\end{eqnarray}
This means that the change in temperature 
of the ordinary matter surrounding 
a mirror matter body
at a distance $R$, $\delta T(R)$,
is given by: 
\begin{eqnarray}
\delta T(R) &=& {-\sigma T'^4 R'^2 \over \kappa R}
\nonumber \\
&\approx &
\left({T' \over 270 \ {\rm K}}\right)^4
\left( {R' \over 5 \ {\rm cm}}\right)^2\left({50 \ {\rm cm} \over R}\right)
\ {\rm K},
\label{late}
\end{eqnarray}
where we have taken
$\kappa = 0.004$ cal/K.s.cm, which is the average value
in the Earth's crust.
[Of course the actual value of $\kappa$ is the one 
valid at the particular impact site].
Clearly, an important but non-trivial solid state
problem is to figure out the rate at which heat can be
transferred from the surrounding ordinary matter into the 
mirror matter object, i.e. what is $T'$? 
If there were perfect thermal conduction between the mirror
matter and surrounding ordinary matter, then $T' = T(R')$.
However, because the thermal conduction is not perfect
(and must go to zero as $\epsilon \to 0$),
it follows that $T' < T(R')$. 
It is possible that $T'$ could be significantly less
than $T(R')$; more work needs to be done to find out.
However, if $T'$ is not so small (it should be
largest for mirror matter bodies with $\epsilon < 0$
and large mirror thermal 
conductivity, such as mirror iron or nickel, because they
can draw in heat from the surrounding ordinary matter throughout
their volume) then Eq.(\ref{late}) 
suggests that mirror matter fragments
can leave a significant imprint on the temperature profile of
the surrounding ordinary matter.
Thus, we may be able to infer the presence of mirror matter
in the ground simply by measuring the temperature of the
Earth at various depths\footnote{
It may be worth looking at satellite infrared temperature maps
of the Tunguska region to see if there is any discernable temperature 
anomaly.}
.

%%%%%%%%%%%%%%%%%%%

\section{Conclusion}

\vskip 0.5cm

One of the most fascinating ideas
coming from particle physics is the concept of mirror matter.
Mirror symmetry -- perhaps the most natural candidate
for a symmetry of nature -- requires a new form of
matter, called `mirror matter', to exist.
The properties of mirror matter make it an ideal
candidate to explain the inferred dark matter of
the Universe. In addition, the mirror matter theory
predicts maximal neutrino oscillations and a shorter
effective orthopositronium lifetime -- both effects
which have been seen in experiments.

If mirror matter really does exist, then some amount should
be out there in our solar system. While there is not much room
for a large proportion of mirror matter
in the inner solar system, it is conceivable that numerous
small (asteroid sized) mirror matter space-bodies might
exist. In fact, it is possible that many of the observed
fireballs are in fact caused by the entry into the Earth's
atmosphere of such mirror matter space-bodies
\footnote{
Interestingly, recent studies using the Sloan
digital sky survey data have found\cite{rec,saibal} that
the number of ordinary space-bodies (greater than
1 km in size) seems to be
significantly less ($\sim 3$
times) than expected from crater rates on the moon\cite{shoe}.
While both ordinary and mirror matter space-bodies would leave
craters on the moon (assuming $\epsilon \sim 10^{-6}$), 
mirror space-bodies may be invisible (or very dark) 
if they contain negligible amount of ordinary matter. Thus,
large mirror space-bodies
may have escaped direct observation, but their presence may have
been hinted by the measured lunar crater rate.
}.
We have shown that the interaction of a mirror matter 
space-body with the Earth's atmosphere seems to
provide a very simple explanation
for the Tunguska event as well as the more puzzling low altitude
fireball events (such as the 1994 Spanish event and 2001 Jordan event
discussed in section II).

Our analysis assumes that the photon-mirror photon
kinetic mixing interaction exists, which is supported by
experiments on orthopositronium.
This fundamental interaction provides the mechanism
causing the mirror space-body to release its kinetic energy in the
atmosphere thereby making its effects `observable'.
Thus, one way to test the Tunguska mirror space-body hypothesis is
to repeat the orthopositronium experiments. If mirror
matter really exists and there is a 
significant photon-mirror photon interaction ($\epsilon > 10^{-9}$),
then this must show up if careful and sensitive experiments
on orthopositronium are done. 

A more dramatic way to test the
mirror space-body hypothesis is to start digging!
If these events are due to the impact of a mirror matter
space-body, then an important implication is that mirror matter
should exist on (or in) the ground at these impact sites. We have
argued that the characteristics of mirror matter fragments on the
Earth's surface depend rather crucially on the effective sign of
the photon-mirror photon kinetic mixing parameter, $\epsilon$,
with the evident lack of surface fragments at 
various `impact sites' suggesting that
$\epsilon < 0$.
In this case, mirror matter should exist embedded in the
ground at the various `impact sites' and can be potentially 
extracted.

\vskip 1.2cm \noindent {\bf Acknowledgements} \vskip 0.5cm 
One of us (R.F.) wishes to greatfully acknowledge
very valuable correspondence with Zdenek Ceplecha, 
Luigi Foschini, Saibal Mitra and Andrei Ol'khovatov.
T. L. Yoon is supported by OPRS and
MRS.


\begin{thebibliography}{999}
\bibitem{ceplecha}
This possibility was pointed out to me by
Z. Ceplecha, Private communication.
\bibitem{saibal}
I would like to thank Saibal Mitra for bringing Refs.\cite{rec,shoe}
and their significance for the mirror matter SB hypothesis, to my attention.
\end{thebibliography}
\end{document}